\documentclass[amstex,aps,showpacs,preprint]{revtex4}%
\usepackage{graphicx}
\usepackage{color}
\usepackage{bm}
\usepackage{longtable}
\usepackage{amsmath}
\usepackage{amsfonts}
\usepackage{amssymb}%
\begin{document}

\title{Weak Measurements in Non-Hermitian Systems}
\author{A.\ Matzkin }
\affiliation{Laboratoire de Physique Th\'{e}orique et Mod\'{e}lisation (LPTM),
CNRS Unit\'{e} 8089,  Universit\'{e} de Cergy-Pontoise, 95302 Cergy-Pontoise cedex, France}

\begin{abstract}
``Weak measurements'' -- involving a weak unitary interaction between a quantum
system and a meter followed by a projective measurement -- are
investigated when the system has a non-Hermitian Hamiltonian. We
show in particular how the standard definition of the ``weak value'' of an
observable must be modified. These studies are undertaken in the context of
bound state scattering theory, a non-Hermitian formalism for which the Hilbert
spaces involved are unambiguously defined and the metric operators can be explicitly
computed. Numerical examples are given for a model system.
\end{abstract}

\maketitle

\section{Introduction}

The standard formulation of quantum mechanics requires physical observables to
be mathematically given in terms of Hermitian operators. In the last decade
theories with a non-Hermitian Hamiltonian have been extensively investigated
\cite{review}. The initial momentum was given by work concerning PT-symmetric
Schr\"{o}dinger operators \cite{bender05}.\ It was hoped that the PT-symmetric
Hamiltonians, which are complex but nevertheless possess a real spectrum,
would provide an extension of standard quantum mechanics. It was later argued
however that these non-Hermitian operators could be mapped to Hermitian ones
by a similarity transform \cite{mosta03}.\ Nevertheless the non-Hermitian
framework remains useful.\ Indeed, from a fundamental perspective it opens up
the possibility of doing quantum mechanics with non-standard inner products.
This has practical consequences because many physical systems are naturally
formulated in non-Hermitian terms \cite{vs}.

Scattering systems involving bound states in real potentials are such a case.
While scattering problems in complex potentials have been prominent in the
studies of non-Hermitian Hamiltonians, bound state scattering in short-range
real potentials have been scarcely investigated
\cite{matz-jpa,znojil08,rotter}. In this case, non-hermiticity arises from the boundary
conditions imposed on the scattering functions, ultimately linked to the fact
that the scattering solutions are not the full eigenstates of the exact
Hamiltonian.\ From the physical point of view this exact Hamiltonian exists,
but its eigenstates are unknown in practice, while the bound scattering
solutions are eigenstates of an effective Hamiltonian that is not Hermitian
relative to the standard inner product. Hence in principle one should employ a
biorthogonal basis, or equivalently obtain the metric operator in order to
define the inner product relative to which the effective Hamiltonian becomes self-adjoint.

In this paper, we will focus on the interaction between a bound state
scattering system and a measurement device in a scheme popularly known as
"Weak measurements" (WM). WM, introduced more than 2 decades ago \cite{aav},
have been receiving an increased attention these last 5 years, in particular
as theoretical but also experimental tool
aimed at investigating fundamental problems in quantum mechanics (see
\cite{phystod} and Refs. therein). WM actually involve two steps: the first step
is a weak interaction between the system and a "weak meter", the overall
evolution being unitary. The second step is a standard projective measurement
in which the system (at that point entangled with the weak meter) interacts
with a different measurement device.\ The state of the system is projected to
a final post-measurement state, while the weak meter has picked up a phase
depending on a quantity known as the "weak value" of the weakly measured
observable. In the conjugate variable of the pointer, the phase-shift appears
as as a shift in the probability distribution.\ This shift can be
experimentally measured by obtaining the probability distribution of the weak pointer.

The main issue when considering a weak measurement of a non-Hermitian system
lies in the treatment of the coupling between the system and the weak meter.
Indeed in a standard projective measurement the observed quantity is an
eigenvalue, which is a real quantity not depending on the definition of the
inner product.\ In a weak measurement, the observed quantity is a shift in the
pointer proportional to the weak value, which, as will be seen below, is a
renormalized transition element. It is therefore crucial in order to determine
the weak value, to properly define the inner product and the physical Hilbert space.

We will first briefly introduce weak measurements and give the usual formula
for computing weak values (Sec 2). We will then derive the weak value for
non-Hermitian systems. In order to provide an unambiguous physical basis, this
derivation will be done in the context of bound state scattering theory. We
will thus explain why systems described with this formalism are non-Hermitian
in the `physical' Hilbert space, leading to the definition of a new inner
product and its associated Hilbert space, in which the weak values must be
defined (Sec\ 3). We will then give in Sec 4 examples of weak value
computations for a model bound scattering system. We will see that
non-Hermitian issues must be incorporated explicitly in order to account for
the correct shift in the weak measurement apparatus. Our concluding remarks
will be given in Sec\ 5.

\section{Weak measurements}

A standard quantum measurement, often represented by the projection of a
premeasurement state of the system to an eigenstate of the measured system
observable, actually involves a two-step procedure.\ First a unitary
interaction between the measured system and the measurement apparatus results
in a system-apparatus state entangled in the pointer basis.\ Then in a second
step the entangled state is projected to a final post-measurement state
correlating a unique pointer state with an eigenstate of the measured observable.

A weak measurement of an observable $A$ proceeds differently.
First a weak unitary interaction takes place between the system and the
"weak" meter. The weakness of the interaction results in an entanglement \cite{sudarshan}
in which the different pointer states are nearly identical. Then a standard
quantum measurement of a different observable takes place, resulting in the
usual projection to an eigenstate of this second observable. Since the system
and the weak apparatus were still entangled, the projection to a final state
of the system also determines the quantum state of the weak meter.

Rather than solving for the weak interaction in terms of the entangled states
in the pointer basis, the standard approach \cite{aav} to weak measurements
starts from a first order expansion of the interaction Hamiltonian.\ Let
$\left\vert \psi(t_{i})\right\rangle $ and $\left\vert \Phi(t_{i}\right\rangle
$ be the initial states of the system and weak meter respectively just before
they interact, and let us assume an interaction Hamiltonian of the form
$I(t)=f(t)A\mathcal{X}$ where the system observable $A$ is coupled to the weak
pointer's position variable along the $x$ axis (ie $\mathcal{X}\left\vert
X\right\rangle =X\left\vert X\right\rangle $ for the pointer). $f(t)$ is a
smooth function of $t$ vanishing outside the interval $t_{-}<t<t_{+}$ during
which the interaction takes place and obeying $\int_{t_{-}}^{t_{+}}f(t)dt=g$ where
$g$ is the mean effective coupling strength.\ Neglecting the self-evolution of
the system and meter during the time interval $t_{+}-t_{-}$, the unitary
evolution generated by $I(t)$ brings the initial state $\left\vert \Psi
(t_{i})\right\rangle \equiv\left\vert \psi(t_{i})\right\rangle \left\vert
\Phi(t_{i})\right\rangle $ to%
\begin{equation}
\left\vert \Psi(t_{+})\right\rangle =e^{-igA\mathcal{X}}\left\vert \psi
(t_{i})\right\rangle \left\vert \Phi(t_{i})\right\rangle . \label{w2}%
\end{equation}
A projective measurement of another observable $B$ of the system is made
immediately after.\ The system state is projected to one of the eigenstates of
$B;$ among the possible outcomes we select only the cases in which the final
state is $\left\vert \beta_{f}\right\rangle .\ $The standard approach consists
in expanding the exponential to first order in $g$ and then compute the
projection
\begin{equation}
\left\langle X\right\vert \left\langle \beta_{f}\right\vert \left.  \Psi
(t_{+})\right\rangle \simeq\left\langle \beta_{f}\right\vert \left.
\psi(t_{i})\right\rangle \left\langle X\right\vert \exp\left(  -ig\frac
{\left\langle \beta_{f}\right\vert A\left\vert \psi(t_{i})\right\rangle
}{\left\langle \beta_{f}\right\vert \left.  \psi(t_{i})\right\rangle
}\mathcal{X}\right)  \left\vert \Phi(t_{i})\right\rangle \label{w5}%
\end{equation}
The term%
\begin{equation}
\left\langle A\right\rangle _{W}\equiv\frac{\left\langle \beta_{f}\right\vert
A\left\vert \psi(t_{i})\right\rangle }{\left\langle \beta_{f}\right\vert
\left.  \psi(t_{i})\right\rangle } \label{w7}%
\end{equation}
is known as the weak value of $A.$ Eq. (\ref{w5}) indicates that the weak
meter has picked up a phase (in configuration space), or alternatively a shift
(in momentum space) proportional to the weak value of $A$, given the initial
(known as "preselected") state, and the final ("postselected") state obtained
after having made a standard measurement of another observable $B$. Note that
the weak value can be a complex number, implying different shifts can be
observed in the conjugate variables of the meter \cite{complex}.

The derivation of Eq. (\ref{w5}) involves several approximations (see eg
\cite{sudarshan}) that will not be discussed here. A necessary (but not
sufficient) condition is that $g\left\langle A\right\rangle _{W}$ is small,
generally implying that the coupling $g$ must be vanishingly small (because
$\left\langle A\right\rangle _{W}$ is generally large).
The important point, from
a physical perspective, is that measuring the weak meter wavefunction allows
to obtain information, encoded in $\left\langle A\right\rangle _{W}$, on the
system observable $A$ without making a full quantum measurement of that
observable.\ Instead another, possibly incompatible property $B$  is measured.\

The applications and interpretations of WM are out of the scope of this work.
Our focus here lies in the weak measurement of a system described in a
non-hermitian framework.\ Indeed, accounting for WM involves treating
-- though only to first order -- the interaction between the non-hermitian
system and a measurement device. Contrary to a standard measurement, in which
case the outcome would be an eigenvalue, the measurement device is shifted by
the weak value. However the definition (\ref{w7}) of the
weak value is valid in standard (Hermitian) quantum mechanics.
For a system described in
a quasi-Hermitian framework
$\left\langle A\right\rangle _{W}$ must be computed in the correct Hilbert space,
endowed with a non-standard inner product, as will be seen below.

\section{Non-Hermitian formalism}

\subsection{General remarks}

The weak value as given by Eq (\ref{w7}) needs to be modified for systems
described in a non-Hermitian setting. The rationale, well-known to
practitioners of PT-symmetric/quasi-Hermitian quantum mechanics, is that the
inner product needs to be replaced. Given the controversies surrounding the
physical interpretation of non-Hermitian systems \cite{conceptual}, our
approach will consist in working with a system -- or rather family of systems,
those involving bound state scattering -- that has an important advantage: the
non-Hermitian aspects appear because one is led to work with wavefunctions
defined on a modified configuration space. This means that while the
scattering Hamiltonian is non-Hermitian, there is in principle an underlying
exact Hamiltonian (though untractable in practice).\ As a result there is no
ambiguity when delaing with the conceptual aspects surrounding non-Hermiticity.

We will therefore first give a brief presentation of bound state scattering,
exposed previously in Ref \cite{matz-jpa}, and discuss its non-Hermitian
aspects in order to derive the formula for the weak value in non-Hermitian
systems, given by Eq (\ref{wv}) below.

\subsection{Bound-state scattering}

For definiteness, let us consider 2\ particles, a light particle and a massive
compound target, attracted by a long-range radial field. The scattering
between the particles is described by a short-range potential. Letting $H^{e}$
denote the exact Hamiltonian in the center of mass, we assume $H^{e}$ can be
split as%
\begin{equation}
H^{e}=H_{0}+V \label{1}%
\end{equation}
where $H_{0}$ is the Hamiltonian of the light particle in the long-rang field
while $V$ contains all the \emph{short range} interactions between the light
particle and the target. We further assume that%
\begin{equation}
\left\langle r^{\prime}\right\vert V\left\vert r\right\rangle =\theta
(r_{0}-r^{\prime})V\theta(r_{0}-r), \label{3}%
\end{equation}
ie $V$ vanishes outside some small radius $r_{0}$ ($\theta$ is the step
function). The total energy $E$ can be partitioned as%
\begin{equation}
E=\varepsilon_{i}+\epsilon_{i}%
\end{equation}
where $\varepsilon_{i}$ is the internal energy of the target (depending on the
target quantum state) and $\epsilon_{i}$ is the energy of the light particle.
The eigenstates of $H_{0}$ are given by
\begin{equation}
\left\vert \phi_{i}(E)\right\rangle =\left\vert f_{i}(\epsilon_{i}%
)\right\rangle \left\vert i(\varepsilon_{i})\right\rangle ; \label{4}%
\end{equation}
$f_{i}(\epsilon_{i},r)\equiv\left\langle r\right.  \left\vert f_{i}%
(\epsilon_{i})\right\rangle $ is the eigenfunction of the radial part of
$H_{0}$ whereas the `target' state $\left\vert i(\varepsilon_{i})\right\rangle
$ includes all the other degrees of freedom, including the non-radial ones of
the colliding particle (a handy notation given that the angular momenta of the
particles are usually coupled). The target states are orthogonal,
$\left\langle i\right\vert \left.  j\right\rangle =\delta_{ij}$. For bound
states $f_{i}(\epsilon_{i},r)$ vanishes at $0$ and $+\infty$ (whenever $E$ is
an eigenvalue of $H_{0}$).

The label $i$ defines the scattering channel. In each channel the
standing-wave solutions are given by the Lippmann-Schwinger equations of
scattering theory as
\begin{equation}
\left\vert \psi_{i}^{e}(E)\right\rangle =\left\vert \phi_{i}(E)\right\rangle
+G_{0}(E)K(E)\left\vert \phi_{i}(E)\right\rangle \label{z50}%
\end{equation}
where $G_{0}(E)$ is the principal-value Green's function and $K$ the reaction
(scattering) operator for standing waves linked to the familiar $S$ matrix by
a Cayley transform \cite{newton82}. The difference here with standard
scattering theory is that the bound channels are included explicitly
\footnote{This means that $G_{0}(E)$ is modified relative to
the usual resolvent by including a term canceling the poles at the
eigenvalues of $H^{e}$
\cite{fano78}.}. Both $\left\langle r\right\vert \left.  \phi
_{i}(E)\right\rangle $ and $\left\langle r\right\vert \left.  \psi_{i}%
^{e}(E)\right\rangle $ diverge as $r\rightarrow\infty$ for an arbitrary value
of $E$.\ A bound state appears when the superposition%
\begin{equation}
\left\vert \psi^{e}(E)\right\rangle =\sum_{i}Z_{i}(E)\left\vert \psi_{i}%
^{e}(E)\right\rangle \label{z52}%
\end{equation}
converges as $r\rightarrow\infty$.\ This happens for discrete values of the energy
obtained, along with the expansion coefficients $Z_{i}(E),$ by imposing the
boundary conditions.

While $H^{e}$ is undoubtedly Hermitian relative to the standard inner product,
its eigenfunctions cannot be computed from Eqs (\ref{z50})-(\ref{z52}) because
the formal expansion of $G_{0}$ over the eigenstates of $H_{0} $ is
intractable. Instead the scattering formulation consists in obtaining a closed
form expression of $G_{0}$ but valid only outside the reaction zone, ie for
$r>r_{0}$. Indeed from the scattering viewpoint, whatever happens within the
reaction zone is encoded in the phase-shifts. The wavefunction (\ref{z52})
outside the reaction zone takes the form%
\begin{equation}
\left\langle r\right.  \left\vert \psi(E)\right\rangle =\sum_{i}Z_{i}(E)
\left[  f_{i}(\epsilon_{i},r)\left\vert i\right\rangle +\sum_{j}g_{j}%
(\epsilon_{j},r)\left\vert j\right\rangle K_{ji}\right]  \qquad r>r_{0}
\label{e20}%
\end{equation}
where $K_{ji}$ are the on-shell elements of the scattering matrix, which are
assumed to be known. $g(r)$ is, like $f(r)$ introduced in eq (\ref{4}), a
solution of the radial part of $H_{0}$ but it is irregular at the origin.

The scattering state $\left\vert \psi(E)\right\rangle $ of (\ref{e20}) is the
part for $r>r_{0}$ of the exact solution $\left\vert \psi^{e}(E)\right\rangle
$, and \emph{not} an approximation to it. But within the scattering
formulation the 'inner' part of $\left\vert \psi^{e}(E)\right\rangle $ for
$r<r_{0}$ does not exist: all meaningful quantities are defined radially on
$[r_{0},\infty\lbrack$. As a consequence,%
\begin{equation}
\left\langle \psi(E_{1})\right\vert \left.  \psi(E_{2})\right\rangle
=\delta_{E_{1}E_{2}}+\mu_{E_{1}E_{2}}(1-\delta_{E_{1}E_{2}}) \label{23b}%
\end{equation}
ie the scattering states are normalized to 1, but are not orthogonal; this is
due to the fact that the boundary conditions at $r=r_{0}$ are not identical
for all the $\left\vert \psi(E)\right\rangle $ \cite{matz-jpa}. Hence the
scattering states cannot be eigenstates of the Hermitian operator
\begin{equation}
H\equiv\sum_{E}E\left\vert \psi(E)\right\rangle \left\langle \psi
(E)\right\vert \label{24}%
\end{equation}
since $H\left\vert \psi(E)\right\rangle \neq E\left\vert \psi(E)\right\rangle
.$ A non-Hermitian Hamiltonian needs to be introduced instead.

\subsection{Non-Hermitian aspects: metric, Hilbert spaces and operators}

Let $\mathcal{H}_{ph}$ be the Hilbert space of standard quantum mechanics.
Physical states are represented by vectors in $\mathcal{H}_{ph}$. From a
practical viewpoint, we may consider that the phase-shifts (or the $K$ matrix
elements) are known and the problem concerns the expansion of physical states
in terms of the scattering solutions $\left\vert \psi(E)\right\rangle .$

To this end let us introduce a non-Hermitian operator $\widetilde{H}$ \ and
state vectors $\left\vert \widetilde{\psi}(E)\right\rangle $ such that%
\begin{align}
&  \widetilde{H}\left\vert \psi(E)\right\rangle =E\left\vert \psi
(E)\right\rangle \label{e31}\\
&  \widetilde{H}^{+}\left\vert \widetilde{\psi}(E)\right\rangle =E\left\vert
\widetilde{\psi}(E)\right\rangle \label{e32}\\
&  \left\langle \widetilde{\psi}(E)\right.  \left\vert \psi(E^{\prime
})\right\rangle =\delta_{EE^{\prime}} \label{e33}%
\end{align}
where $\{\left\vert \widetilde{\psi}(E)\right\rangle ,\left\vert
\psi(E)\right\rangle \}$ forms a biorthogonal basis. It follows that we can
write the following expansions:%
\begin{equation}
\widetilde{H}=\sum_{E}E\left\vert \psi(E)\right\rangle \left\langle
\widetilde{\psi}(E)\right\vert \qquad\widetilde{H}^{+}=\sum_{E}E\left\vert
\widetilde{\psi}(E)\right\rangle \left\langle \psi(E)\right\vert . \label{e34}%
\end{equation}
$\widetilde{H}$ and $\widetilde{H}^{+}$ are further linked by%
\begin{equation}
\widetilde{H}=\mathcal{G}\widetilde{H}^{+}\mathcal{G}^{-1} \label{e40}%
\end{equation}
where $\mathcal{G}$ is a Hermitian operator given by%
\begin{equation}
\mathcal{G}=\sum_{E}\left\vert \psi(E)\right\rangle \left\langle
\psi(E)\right\vert \qquad\mathcal{G}^{-1}=\sum_{E}\left\vert \widetilde{\psi
}(E)\right\rangle \left\langle \widetilde{\psi}(E)\right\vert . \label{e41}%
\end{equation}
Eq (\ref{e40}) is the defining relation of quasi-Hermiticity
\cite{mostafazadeh batal04} $\mathcal{G}$ being invertible and
positive-definite \cite{matz-jpa}.

This allows to define a Hilbert space $\mathcal{H}$ endowed with a new inner
product depending on the metric $\mathcal{G}$:%
\begin{equation}
\left(  \psi(E_{1}),\psi(E_{2})\right)  _{\mathcal{G}}\equiv\left\langle
\psi(E_{1})\right\vert \mathcal{G}^{-1}\left\vert \psi(E_{2})\right\rangle
=\left\langle \widetilde{\psi}(E_{1})\right.  \left\vert \psi(E_{2}%
)\right\rangle =\delta_{E_{1}E_{2}}. \label{e45}%
\end{equation}
Eq (\ref{e40}) indicates that $\widetilde{H}$ is Hermitian relative to this
new inner product. Completeness of the biorthogonal basis allows to expand an
arbitrary state of $\mathcal{H}_{ph}$ in terms of the $\left\vert
\psi(E)\right\rangle $, ie the eigenstates of $\widetilde{H}$ span the entire
Hilbert space of admissible physical states even if they do not form an
orthogonal basis in $\mathcal{H}_{ph}$ \footnote{We will not address here the
delicate technical aspects related to completeness in the case of infinite
dimensional Hilbert spaces.}

Calculations involving the scattering states have to be performed in
$\mathcal{H}$ rather than in $\mathcal{H}_{ph}$. Indeed although a physical
state $\left\vert \alpha_{k}\right\rangle _{ph}$ is known in $\mathcal{H}%
_{ph}$, its expansion over the scattering eigenstates $\left\vert \alpha
_{k}\right\rangle =\sum_{E}a_{k}(E)\left\vert \psi(E)\right\rangle $ is
defined in $\mathcal{H}$ with the expansion coefficients given through%
\begin{equation}
a_{k}(E)=\left\langle \widetilde{\psi}(E)\right.  \left\vert \alpha
_{k}\right\rangle \equiv\left(  \psi(E),\alpha_{k}\right)  _{\mathcal{G}%
}=\left\langle \psi(E)\right.  \left\vert \widetilde{\alpha}_{k}\right\rangle
\label{14}%
\end{equation}
where we have put $\left\vert \widetilde{\alpha}_{k}\right\rangle
\equiv\mathcal{G}^{-1}\left\vert \alpha_{k}\right\rangle .$ Note that in the
underlying \emph{exact} problem, there is a physical state corresponding to
$\left\vert \alpha_{k}\right\rangle $ and given by the same expansion
coefficients but over the eigenstates of the exact Hamiltonian, $\left\vert
\alpha_{k}^{e}\right\rangle =\sum_{E}a_{k}(E)\left\vert \psi^{e}%
(E)\right\rangle $ with $a_{k}(E)=\left\langle \psi^{e}(E)\right.  \left\vert
\alpha_{k}^{e}\right\rangle .$ We can therefore understand non-Hermiticity as
a consequence of working with exact wavefunctions but defined only over part
of configuration space relative to the underlying exact problem.

As was the case with $H$ [Eq (\ref{24})] that needed to be replaced with
$\widetilde{H}$, an operator $A$ Hermitian in $\mathcal{H}_{ph}$ is
represented in $\mathcal{H}$ by an operator $\widetilde{A}$ whose expansion
over the biorthogonal basis reads
\begin{equation}
\widetilde{A}=\sum_{EE^{\prime}}\left\vert \psi(E)\right\rangle \widetilde
{A}_{EE^{\prime}}\left\langle \widetilde{\psi}(E^{\prime})\right\vert .
\label{20}%
\end{equation}
The relation between $A$ and $\widetilde{A}$ is given by \cite{mostafazadeh
batal04,matz-jpa} $A=\mathcal{G}^{-1/2}\widetilde{A}\mathcal{G}^{1/2}$. The
time evolution operator is a prominent example: $U(t)=\sum_{E}e^{-iEt}%
\left\vert \psi(E)\right\rangle \left\langle \psi(E)\right\vert $ is
\emph{not} unitary in $\mathcal{H}_{ph}.\ $The correct unitary operator in
$\mathcal{H}_{ph}$ is obtained from $U=\mathcal{G}^{-1/2}\widetilde
{U}\mathcal{G}^{1/2}$ where $\widetilde{U}(t)$ defined by%
\begin{equation}
\widetilde{U}(t)=\sum_{E}e^{-iEt}\left\vert \psi(E)\right\rangle \left\langle
\widetilde{\psi}(E)\right\vert \label{e50}%
\end{equation}
is (pseudo) unitary in $\mathcal{H}$.

\subsection{Weak values}

We are now in a position to formulate the weak value expression for a
non-Hermitian system. First note that, as described in Sec. 2, we
do not need to solve explicitly the full problem involving the coupling
of a non-Hermitian system to a Hermitian one (as eg in Ref. \cite{coupling}).
Indeed the non-Hermitian system is practically not affected by the weak
interaction, while the Hermitian one (the meter) simply picks up a
phase. This phase -- the weak value -- is therefore the only quantity we need
to determine.

Let $\left\vert \alpha_{i}\right\rangle =\sum_{E}%
a_{i}(E)\left\vert \psi(E)\right\rangle $ be the initial ("preselected") state
prior to the weak measurement of a system observable $A $ and $\left\vert
\beta_{f}\right\rangle =\sum_{E}b_{f}(E)\left\vert \psi(E)\right\rangle $ the
"postselected" state obtained after the subsequent projective measurement.
According to the discussion above, $\left\vert \alpha_{i}\right\rangle $ and
$\left\vert \beta_{f}\right\rangle $ represent the physical states in
$\mathcal{H}$ and the coefficients $a_{i}$ and $b_{f}$ are given by formulae
analog to Eq (\ref{14}).\ The observable $A $ is represented in $\mathcal{H}$
by the non-Hermitian operator $\widetilde{A}$ whose expansion over the
biorthogonal basis was given by Eq (\ref{20}).

The formula (\ref{w7}) expressing the weak value of the observable $A$ becomes%
\begin{equation}
\left\langle \widetilde{A}\right\rangle _{W}=\frac{(\beta_{f},\widetilde{A}\alpha
_{i})_{\mathcal{G}}}{(\beta_{f},\alpha_{i})_{\mathcal{G}}}%
\end{equation}
or in terms of the standard inner product notation%
\begin{equation}
\left\langle \widetilde{A}\right\rangle _{W}=\frac{\left\langle \widetilde
{\beta_{f}}\right\vert \widetilde{A}\left\vert \alpha_{i}\right\rangle
}{\left\langle \widetilde{\beta_{f}}\right\vert \left.  \alpha_{i}%
\right\rangle }=\frac{\left\langle \beta_{f}\right\vert \mathcal{G}%
^{-1}\widetilde{A}\left\vert \alpha_{i}\right\rangle }{\left\langle \beta
_{f}\right\vert \mathcal{G}^{-1}\left\vert \alpha_{i}\right\rangle }%
=\frac{\left\langle \beta_{f}\right\vert \mathcal{G}^{-1/2}A\mathcal{G}%
^{-1/2}\left\vert \alpha_{i}\right\rangle }{\left\langle \beta_{f}\right\vert
\mathcal{G}^{-1}\left\vert \alpha_{i}\right\rangle }. \label{wv}%
\end{equation}
Hence when a system that is non-Herrmitian (relative to the standard inner
product) interacts with a weak measurement apparatus measuring the observable
$A$, the meter is shifted by a quantity given by Eq (\ref{wv}), not by
Eq\ (\ref{w7}); Eqs (\ref{wv}) and \ (\ref{w7}) obviously coincide when the
metric is flat ($\mathcal{G}$ is the identity operator). Note that the shift
in the weak measurement apparatus can in principle be experimentally observed.

\section{Computation of weak values in a model non-hermitian system}

\subsection{Model}

We will give examples involving the computation of Eq (\ref{wv}) in a
situation well-known in atomic physics involving atoms with a single excited
electron. In this case the long-range field is the familiar Coulomb potential
and the reaction zone is about the size of the atomic core. The excited
electron periodically scatters off the core, the core-electron interaction
being embodied in the short-range potential.\ We set up a model with 5
scattering channels: the target has a ground state with $\varepsilon_{1}=0$
and 4 excited states with energies $\varepsilon_{i},i=2,...,5$.\ The
$5\times5$ scattering matrix $K(E)$ is chosen to have a very strong energy
dependence \footnote{If $K$, or equivalently the phase-shifts, are taken as
independent of the energy, then the problem remains formally non-Hermitian,
but the non-Hermitian character becomes negligible in practical computations
\cite{matz-jpa} as $\mathcal{G}$ is then nearly identical to the identity
(flat metric).} in order to have stronger non-diagonal elements of the metric
$\mathcal{G}$ (a non-Hermiticity index $\kappa$ can be defined by averaging
over the $N$ largest non-diagonal elements of $\mathcal{G}$ where $N$ is the
dimension of the metric; in this model we have $\kappa=0.061$, a non-negligible
though relatively samll value). The bound state energies $E$ are obtained
numerically by enforcing the boundary conditions in Eq (\ref{e20}) and then
the coefficients $Z_{i}(E)$ are retrieved by solving the relevant linear
system. While the number of bound states is infinite, good numerical
convergence is obtained by taking about 200 states above and 200 states below
the energy interval of interest. The metric employed in the numerical
computations is thus a $400\times400$ matrix.

For the purpose of illustration we will determine the weak value of the radial
position of the excited electron and the weak value of the energy, assuming in
both cases postselection can be made to a final state identical to the initial
one. We choose an initial state $\left\vert \alpha(t=0)\right\rangle $, that
we take to be a Gaussian localized radially very far from the target, at the
outer turning point of the radial potential for an excited electron (with a
mean energy $n=42$), with the target being in its ground state. Initially
$\left\vert \alpha(t=0)\right\rangle $ is defined on an orthogonal basis of
$\mathcal{H}_{ph}$ but we assume (and verify numerically) that this state can
approximately be expanded on our chunk of computed eigenstates of
$\widetilde{H}$ as%
\begin{equation}
\left\vert \alpha(t=0)\right\rangle =\left\vert F_{loc}(r\approx
r_{tp})\right\rangle \left\vert \varepsilon_{1}\right\rangle =\sum
a(E)\left\vert \psi(E)\right\rangle \label{init}%
\end{equation}
where the $a(E)$ are determined as in Eq (\ref{14}). We now proceed to
compute weak values.

\subsection{Weak value of the energy}

We consider a scheme in which a weak measurement of the energy is made at
$t=t_{W}$, immediately followed by a projection to a final state. We assume
for definiteness it is possible to postselect on a state $\left\vert \beta
_{f}\right\rangle $ identical to the initial state $\left\vert \alpha
(t=0)\right\rangle $, for example by considering a weak measurement apparatus
consisting in an array of devices placed spherically at a radial distance
$r\approx r_{tp}$ from the atomic core; the weak measurement time $t_{W}$ must
then correspond to the recurrence time (when the wavepacket relocalizes periodically at the turning point
\cite{nauenberg}) in the initial scattering channel, here
channel 1.

If non-Hermitian issues are ignored, then the operator $H$ of Eq (\ref{24})
would be employed for the Hamiltonian, the evolution operator, accounting for
the evolution of the system form $t=0$ to $t_{W}$ would be taken as
$U(t)=\sum_{E}e^{-iEt}\left\vert \psi(E)\right\rangle \left\langle
\psi(E)\right\vert $ and the weak value obtained from the usual definition
(\ref{w7}) would thus be given by%
\begin{align}
\left\langle H(t_{W})\right\rangle _{W}\text{\textquotedblleft}  &
=\text{\textquotedblright}\frac{\left\langle \beta_{f}\right\vert H\left\vert
\psi(t_{W})\right\rangle }{\left\langle \beta_{f}\right\vert \left.
\psi(t_{W})\right\rangle }\label{w10}\\
\text{\textquotedblleft}  &  =\text{\textquotedblright}\frac{\left\langle
\alpha(t=0)\right\vert HU(t_{W})\left\vert \alpha(t=0)\right\rangle
}{\left\langle \alpha(t=0)\right\vert U(t_{W})\left\vert \alpha
(t=0)\right\rangle }. \label{w12}%
\end{align}
This quantity is plotted in Fig.\ 1 (dotted lines) for different possible
choices of the measurement time $t_{W}$ compatible with the system wavepacket
radially localized in the neighborhood of the measuring apparatus.

However, since the system \ is non-Hermitian, Eqs (\ref{w10})-(\ref{w12})
should formally be replaced by%
\begin{align}
\left\langle \widetilde{H}(t_{W})\right\rangle _{W}  &  =\frac{\left\langle
\widetilde{\beta_{f}}\right\vert \widetilde{H}\left\vert \psi(t_{W}%
)\right\rangle }{\left\langle \widetilde{\beta_{f}}\right\vert \left.
\psi(t_{W})\right\rangle }\label{w13}\\
&  =\frac{\left\langle \widetilde{\alpha}(t=0)\right\vert \widetilde
{H}\widetilde{U}(t_{W})\left\vert \alpha(t=0)\right\rangle }{\left\langle
\widetilde{\alpha}(t=0)\right\vert \widetilde{U}(t_{W})\left\vert
\alpha(t=0)\right\rangle } \label{w15}%
\end{align}
where $\widetilde{H}$ is the non-Hermitian Hamiltonian given by Eq (\ref{e34})
and $\widetilde{U}(t)$ is the corresponding evolution operator given by Eq
(\ref{e50}). This quantity is also plotted in Fig 1 (solid line).

The results shown in Fig.\ 1 indicate a similar overall behaviour for the two
curves, though there are substantial differences for several values of the
measurement time \footnote{Fig.1 shows the real part of the weak value --
there is also an imaginary part that is several orders of magnitude smaller.}
Therefore the replacement of the usual formulae (\ref{w10})-(\ref{w12}) by
Eqs. (\ref{w13})-(\ref{w15}) is not purely formal: in practical computations
the non-Hermitian nature of the system, coupled to a weak measurement device,
\emph{must} be taken into account in order to compute correctly the expected
shift in the pointer of the measurement apparatus due to the weak measurement.

\begin{figure}[tb]
\begin{center}
\includegraphics[height=2.1in,width=3in]{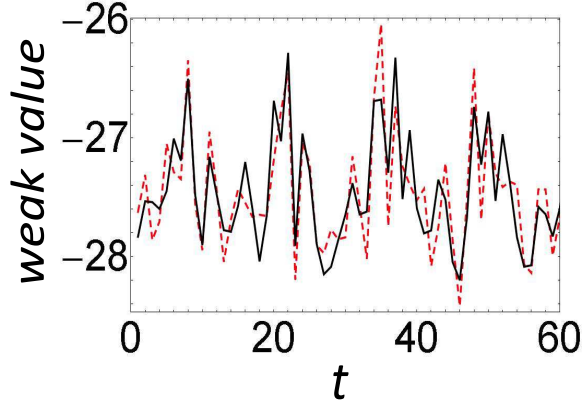}
\end{center}
\caption{The weak value of the energy (in atomic units$/10^{-5}$) is
shown for different measurement times compatible with the system wavepacket
being in the neighborhood of the weak measurement device. The dashed (gray,
online red) line represents the usual weak value expression, given by Eqs
(\ref{w10})-(\ref{w12}). The solid black line represents the correct
expressions Eqs (\ref{w13})-(\ref{w15}) for the weak value of the energy
in non-Hermitian systems. The time is given in units of the wavepacket period
(about $1.15\times10^{-11}$ s). The average energy of the system is
$27.89\times10^{-5}$ au. }%
\label{fig1}%
\end{figure}

\subsection{Weak value of the momentum}

Another example is the weak value of the momentum postselected to a given
position.\ This has become a standard example \cite{leavens} involving a weak
measurement of an observable which is incompatible with the postselected one.
Here, rather than postselecting to a position $\left\vert r\right\rangle $
known with an infinite precision, we employ as above $\left\vert \beta
_{f}\right\rangle =\left\vert \alpha(t=0)\right\rangle $ as the postselected
state, keeping in mind that $\left\langle r\right\vert \left.  \alpha
(t=0)\right\rangle $ is a wavefunction tightly localized around the turning
point $r_{tp}$ [Eq (\ref{init})]. As in the previous example we assume the
weak measurement on the preselected state can be made at different times
$t_{W}$ for which the system wavepacket relocalizes in the neighborhood of the
measuring apparatus.

\begin{figure}[tb]
\begin{center}
\includegraphics[height=2.1in,width=3in]{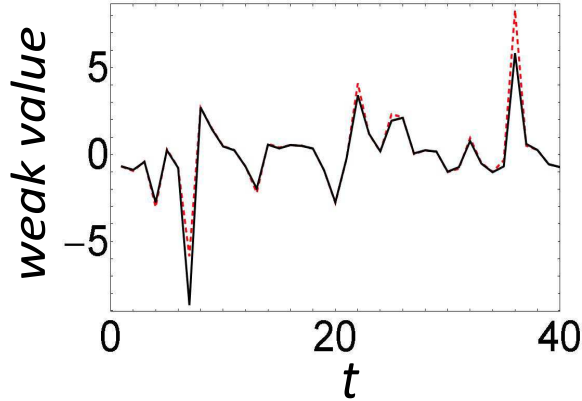}
\end{center}
\caption{The weak value of the momentum (in atomic units$/10^{-3}$) is
shown for different measurement times compatible with the system wavepacket
being in the neighborhood of the weak measurement device. The dashed (gray,
online red) line represents the usual weak value expression while
the solid black line shows the expression valid for non-Hermitian systems.
The time is given in units of the wavepacket period
(about $1.15\times10^{-11}$ s).  }%
\label{fig1}%
\end{figure}

The weak value (\ref{wv}) becomes%
\begin{equation}
\left\langle \widetilde{P}(t_{W})\right\rangle _{W}=\frac{\left\langle
\widetilde{\beta_{f}}\right\vert \widetilde{P}\left\vert \psi(t_{W}%
)\right\rangle }{\left\langle \widetilde{\beta_{f}}\right\vert \left.
\psi(t_{W})\right\rangle }\label{150}%
\end{equation}
with $\left\vert \psi(t_{W})\right\rangle =\widetilde{U}(t_{W})\left\vert
\alpha(t=0)\right\rangle $. Note that $\left\langle \widetilde{P}%
(t_{W})\right\rangle _{W}$ has both a real and a complex part: the real part
is related to the average velocity field of the system while the complex part
is proportional to the logarithmic derivative of the system wavefunction
modulus \cite{jordan}.\ In principle both the real and the complex parts can
be experimentally observed (though not jointly). The real part of the weak
value (\ref{150}) is plotted in Fig.\ 2 (black solid line). The dashed line is
obtained by a straightforward application of Eq (\ref{w7}), ie when the
non-Hermitian character of the system is not taken into account. The two
curves nearly overlap, which can appear as a little surprising in view of the
fact that $U(t)$ is not unitary and therefore probability is not
conserved.\ Notwithstanding there are measurement times for which the
discrepancy between Eq (\ref{w7}) and the correct Eq (\ref{wv}) is important.

\section{Summary and Conclusion}

We have investigated weak measurements for quantum systems described by a
non-Hermitian Hamiltonian. The standard definition (\ref{w7}) of the weak
value -- the quantity that can in principle be experimentally observed by
reading the pointer of a weak meter --  does not hold in a non-Hermitian
framework. The modified expression, given by Eq (\ref{wv}) was derived in this
work in the context of bound state scattering theory. The advantage of
employing this particular instance of non-Hermitian formalism is that its
physical meaning is devoid of any ambiguity, though the validity of Eq
(\ref{wv}) holds in general (at least when the relevant similarity transform
can be defined).

The results were illustrated numerically in a model system by computing the
weak values of the energy and of the momentum, with a postselection to a state
identical to the initial radially localized wavefunction. Overall, the results
indicate that even in a non-Hermitian in which the non-diagonal elements of
the metric are relatively small (the non-Hermiticity index was $\kappa
=0.061\ll1$), it is important in concrete studies of weak measurements to
employ the correct (ie non-Hermitian) formulae in order to account
appropriately for the behaviour of the weak meters.


\vspace{1cm}

\begin{thebibliography}{99}                                                                                               %


\bibitem {review}Mostafazadeh A 2010 Int J Geom Methods Mod Phys 7 1191

\bibitem {bender05}Bender C M 2005, Contemporary Phys 46 277.



\bibitem {mosta03}Mostafazadeh A 2003, J Phys A 36 7081

\bibitem {vs}Ruschhaupt A, Delgado F and Muga J G 2005 J Phys A 38 L171;

Bender CM, Chen JH and Milton KA 2006 J Phys A 39 1657;

Znojil M 2008 J Phys A 41 215304;

Jones HF and Rivers RJ 2009 Phys Lett A 37 3304

\bibitem {matz-jpa}Matzkin A 2006 J Phys A 39 10859

\bibitem {znojil08}Znojil M 2008 J Phys A 41 292002

\bibitem {rotter}Rotter I 2009 J Phys A 42 153001

\bibitem {aav}Aharonov Y, Albert DZ and Vaidman L 1988 Phys. Rev. Lett. 60, 1351

\bibitem {phystod}Aharonov Y, Popescu S and Tollaksen J 2010 Phys. Today 63, 27

\bibitem {sudarshan}Duck I M, Stevenson P M and Sudarshan E C G 1989 Phys Rev
D 40 2112;

Pan A K and Matzkin A 2012, Phys Rev A 85 022122

\bibitem {complex}Jozsa R 2007 Phys Rev A 76 044103

\bibitem {conceptual}Mostafazadeh A 2010 Phys Scr 82 038110

\bibitem {newton82}Newton R G 1982 \emph{Scattering theory of waves and
particles}, NewYork : Springer.

\bibitem {fano78}Fano U 1978, Phys Rev A 17 93;

Matzkin A 1999, Phys Rev A 59 2043

\bibitem {mostafazadeh batal04}Mostafazadeh A and Batal A 2004, J Phys A 37 11645

\bibitem {coupling}Bender C M and Jones H F 2008 J. Phys. A 41 244006


\bibitem {nauenberg}Suarez Barnes I M, Nauenberg M, Nockleby M and Tomsovic S
1994 J Phys A 27 3299

\bibitem {leavens}Leavens C\ R 2005 Found Phys 35 469

\bibitem {jordan}Dressel J\ and Jordan A N 2012 Phys Rev A 85  012107
\end{thebibliography}
\end{document}